# Electric field manipulation enhanced by strong spin-orbit coupling: promoting rare-earth ions as qubits

Zheng Liu[1], Ye-Xin Wang[1], Yu-Hui Fang[1], Si-Xue Qin[2], Zhe-Ming Wang[1], Shang-Da Jiang[1,3*], and Song Gao[1,3,4*]

1.  Beijing National Laboratory of Molecular Science, State Key Laboratory of Rare Earth Materials Chemistry and Applications, Beijing Key Laboratory of Magnetoelectric Materials and Devices, College of Chemistry and Molecular Engineering, Peking University, Beijing 100871, P. R. China. E-mail: jiangsd@pku.edu.cn; gaosong@pku.edu.cn
2.  Department of Physics, Chongqing University, Chongqing 401331, P. R. China
3.  Beijing Academy of Quantum Information Sciences, West Bld.#3, No.10 XiBeiWang East Rd., HaiDian District, Beijing 100193, P. R. China
4.  School of Chemistry and Chemical Engineering, South China University of Technology, Guangzhou 510640, P. R. China

**Abstract**

Quantum information processing based on magnetic ions are considered potential candidates for applications because they can be modified and scaled up by a variety of chemical methods. For these systems to achieve individual spin addressability and high energy efficiency, we exploited the electric field as a tool to manipulate their quantum behaviours, functioning via spin-orbit coupling. A Ce:YAG single crystal was employed due to that rare-earth ions have strong spin-orbit coupling and with considerations regarding the dynamics and the symmetry requirements. The Stark effect of the $Ce^{3+}$ ion was observed and measured. When demonstrated as a quantum phase gate, the electric field manipulation exhibited high efficiency which allowed up to 57 $\pi/2$ operations before decoherence with optimized field directions. It was also utilized to carry out quantum bang-bang control, as a method of dynamic decoupling, and the refined Deutsch-Jozsa algorithm. Our experiments highlighted rare-earth ions as potentially applicable qubits since they offer enhanced spin-electric coupling which enables high-efficiency quantum manipulation.

**Introduction**

Quantum computation offers accelerated ways of solving problems such as database searching[1] and prime factor decomposition[2]. Recently, there is an uprising trend to employ magnetic molecules as quantum bits (qubits), due to their advantages including monodispersed size and chemically controllable properties[3]. In these magnetic molecules where the electron spin is utilized, the electronic structure and quantum coherence are tunable by modifying the environment of the spin[4], and the dimension of the Hilbert space can be increased by designing the magnetic coupling between the spin carriers. Recently, the quantum phase memory time of the electron spin has been largely increased to nearly a millisecond by molecular design using nuclear spin-free ligands[5], showing a bright future for quantum information processing based on electron spin qubits.

However, the magnetic field is hardly desirable in controlling the electron spins in the quantum information processing context since addressability is difficult to realize due to its poor locality. The electric field (***E*** field), which can be much more concentrated and efficient, offers a promising alternative way. The coupling between the ***E*** field and spins are referred to as spin-electric coupling and observed in the Stark effect. Related magnetic resonance researches were first published in the 1960s[6,7] and later reviewed by Mims[8].

The electric control of spins has been proposed to realize electric-dipole-induced spin resonance in semiconductors[9], and numerous experiments have been performed to control magnetic quantum dots[10], single electron spins[11], diamond defects[12] and the single nuclear spin in a molecular magnet[13]. The coherent control of an electron spin ensemble has been demonstrated in pnictogen-doped silicon semiconductors[14–16], piezo-diluted magnetic semiconductors[17] and molecular magnets[18]. Most recently, Morello has reported nuclear electric resonance using a high-spin $^{123}$Sb nuclear doped in silicon, indicating the importance the electric control of the spin in quantum information processing[19].

The electric control of electron spin was not observed for some of the molecular magnets reported as qubits, and observed to be inefficient in others, especially those with light elements. This is mainly because the spin-electric coupling is weak, and hinders their further application as qubits. This article is dedicated to demonstrate the solving of this problem, using an ionic crystal. We report largely enhanced spin-electric coupling in rare-earth ions, with which the coherent control by the *E* field is illustrated as a phase gate. Using this electric phase gate, quantum bang-bang decoupling is realized with microwave (mw) pulses, and the refined Deutsch-Jozsa (D-J) algorithm is demonstrated.

**Backgrounds: The role of spin-orbit coupling**

The spin and orbital motions of electrons generate magnetic moments and contribute to a variety of spectroscopic and chemical phenomena. Unlike main group elements and transition metals, rare-earth ions have their orbital angular momentum originating from the inner subshell of 4*f*, and thus unquenched by the crystal field, making the spin-orbit coupling dominate the properties of 4*f* electrons. In general, the Hamiltonian of a paramagnetic ion can be written as

$$\hat{H} = \mu_B \vec{B}^T \cdot (\hat{L} + g_e \hat{S}) + \lambda \hat{L}^T \cdot \hat{S} + \hat{H}_{CF} \tag{1}$$

where the three terms represent the Zeeman splitting, the spin-orbit coupling and the crystal field effect, which results basically from the electrostatic field around the ion. The external *E* field Hamiltonian $\hat{H}_E$ can be viewed as a variation to the crystal field term due to changes including atomic displacements and charge redistribution. To the first order it can be expressed as $\hat{H}_E = (\partial \hat{H}_{CF}/\partial E_x, \partial \hat{H}_{CF}/\partial E_y, \partial \hat{H}_{CF}/\partial E_z) \cdot \vec{E}$. In cases where the spin-orbit coupling can be treated as a perturbation, the above Hamiltonian can be converted and expressed in the spin operator only, as

$$\hat{H} = \mu_B \vec{B}^T \cdot \overline{\overline{g}} \cdot \hat{S} + \hat{S}^T \overline{\overline{D}} \hat{S} \tag{2}$$

where $\overline{\overline{g}} = g_e \overline{\overline{1}} + 2\lambda \overline{\overline{\Lambda}}$, $\overline{\overline{D}} = \lambda^2 \overline{\overline{\Lambda}}$, and $\overline{\overline{\Lambda}}$ is a 3×3 matrix containing the orbital angular momentum eigenstates and their energies determined by the crystal field term. The effect of $\hat{H}_E$ is then included via $\overline{\overline{\Lambda}}$ in both the Zeeman and zero-field splitting term and scales up with $\lambda$. Cases with weak crystal field and strong spin-orbit coupling require more complicated theoretical treatments, but similar basic principle still holds, indicating that strong spin-orbit coupling is desirable in seeking materials for efficient *E* field control of a spin system. A recent research concludes that, with similar structures, the spin-orbit coupling is not the primary factor leading to decoherence[20]. This ensures the possibility to employ strong spin-orbit coupling systems as qubits while maintaining long quantum coherence time. We therefore propose that the rare-earth ions are desirable for strong spin-electric coupling.

The Stark effect has been observed in several bulk systems including pnictogen-doped silicon semiconductors[16] and piezoelectric materials[17], where the spin-obit coupling of the spin carriers is weak, but the applied *E* field dramatically redistributes the electron wave function. Differently, in this research, the *E* field manipulates the quantum coherence properties of the spins localized on the rare-earth ions individually.

Spin-orbit coupling may also help via spin-spin coupling. Antisymmetric exchange, also known as the Dzyaloshinskii-Moriya (D-M) interaction, arises from the electron spin anisotropy due to large spin-orbit coupling. This kind of exchange results in spin canting and net electric polarization, which provide strong spin-electric coupling. R. Sessoli and co-workers tuned the magnetic interaction between an organic radical and a $Mn^{2+}$ in a single crystal in a chiral space group[21]. By applying the ac ***E*** field to the sample continuously, the electron paramagnetic resonance (EPR) spectra showed a shift of the effective *g*-factor up to $-2.5\times10^{-6}$.

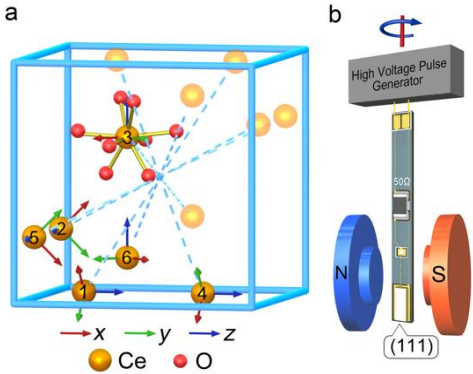

**Figure 1** The Ce:YAG crystal structure and the device used in our experiments. **a)** Locations of the six magnetically inequivalent $Ce^{3+}$ ions marked as **Ce-1** to **Ce-6**. Each $Ce^{3+}$ is coordinated by 8 oxygen atoms. Their principle axes are shown as coloured arrows (red, green and blue for *x*, *y* and *z*, respectively). Inversion centres are located at (0.5, 0.5, 0.5) and (0.5±0.25, 0.5±0.25, 0.5±0.25). The inversion pairs involving **Ce-1** to **Ce-6** generated by (0.5, 0.5, 0.5) are also shown. **b)** Ce:YAG single crystal of size $2\times6\times0.5$ mm$^3$ is mounted on the electrode, and connected to 0.2 mm gold wires with a 50 Ω match resistor.

## Results and discussions

**Observation of the spin-electric coupling**

The Stark effect can be observed with two methods. A straightforward one is to apply the ***E*** field to the sample continuously to modify the energy levels of the spin carrier. The slight influence can be observed with a strong dc ***E*** field or, via signal modulation, with an ac ***E*** field. Another approach, which is more efficient and common for molecular magnets, is to measure via quantum phase evolution with pulsed EPR[18].

Previous researches using pulsed EPR were performed with powder samples. To evaluate the spin-electric coupling in more details, a single crystal was used in our experiments. The emergence of the Stark effect in magnetic resonance requires that the spin carriers must not be located at any inversion centre. On the other hand, the number of inequivalent parameters needed to model the effect would be smaller if the spin centres are of higher symmetry. Based on these considerations, an yttrium aluminium garnet (YAG) single crystal doped with $Ce^{3+}$ was employed. YAG crystallizes in the space group $Ia\bar{3}d$ with cubic unit cells. $Ce^{3+}$ was doped into the YAG crystal at the $Y^{3+}$ positions[22]. There are 6 magnetically inequivalent sites[23] and each $Ce^{3+}$ is identically coordinated by 8 oxygen atoms with a local symmetry of $D_2$. The three $C_2$ axes serve as principal axes for the *g*-factor. $Ce^{3+}$ ions appear in inversion pairs due to the central symmetry of the crystal [Figure 1(a)]. To eliminate the electron spin dipolar interactions, the concentration of $Ce^{3+}$ ions was reduced to less than 0.1%, and the average distance between $Ce^{3+}$ ions was more than 3 nm. In order to enhance the ***E*** field strength and acquire a significant signal, the Ce:YAG single crystal was cut to 0.5 mm in thickness, with a $2\times6$ mm$^2$ area to fit the sample into the cavity. The crystal was mounted so that the (111) face was perpendicular to the E field and parallel with the $B_0$ field. This geometry was fixed for all our experiments.

With strong spin-orbit coupling, the $Ce^{3+}$ with $4f^1$ electron configuration presents a $J = 5/2$ state. Due to the large crystal field splitting, only the ground doublets can be observed in X-band EPR measurements. For simplicity, an effective spin 1/2 model is used, where the values of the highly anisotropic $g$-tensor depend on the crystal field parameters. Hence, the spin-electric coupling, enhanced by large spin-orbit coupling and significantly influencing the crystal field parameters, can be detected through the $g$-factor shift in our experiments.

In this model, the Hamiltonian of the $Ce^{3+}$ ion in the external static magnetic field is $\hat{H} = \mu_B B^T \bar{\bar{g}} \hat{S}$. By cw-EPR measurements and simulations, the principal values of the $g$-tensor are determined to be $g_{xx}$=1.85, $g_{yy}$=0.90 and $g_{zz}$=2.74 which agree with the previous report[23].

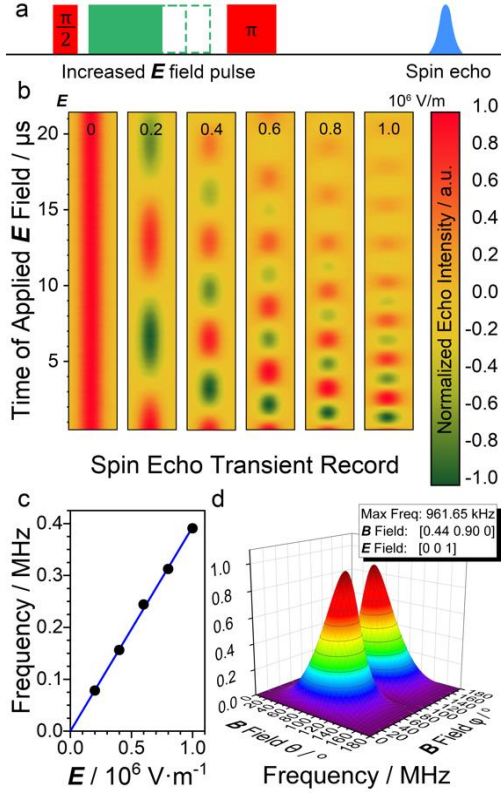

**Figure 2** Spin echo oscillation upon the application of an $E$ field. **a)** The pulse sequence employed in the experiment. The rectangular $E$ field pulse is applied between the mw $\pi/2$ and $\pi$ pulses of the standard Hahn echo pulse sequence. **b)** The transient spin echo oscillates with increasing frequencies upon enhancing the $E$ field strength from 0 to $10^6$ V/m. **c)** The quantum phase evolution frequency, derived from the Fourier transformation of (b), shows a linear dependence on the $E$ field strength. The error bars lie within the size of the points. **d)** The quantum phase evolution frequency is simulated with varying directions of the $E$ and $B_0$ fields. The most efficient frequency for quantum phase control is optimized to be 0.96 MHz when $E \parallel z$ and $B_0 \parallel$ (0.44, 0.90, 0) with respect to the local $Ce^{3+}$ coordinates. The experiment was conducted at 10 K, with the microwave frequency 9.667 GHz.

We applied the $E$ field pulse between the two mw pulses in the standard Hahn echo sequence. An oscillation of the spin echo can be observed when the duration of the $E$ field pulse is increased, indicating a continuously changing quantum phase. This phase evolution can be used to characterize the Stark effect. The effect of the $E$ field is mainly modifying the Zeeman splitting term, so we can generally express the Stark effect Hamiltonian as

$$\hat{H}_E = \sum_{ijk} E_i T_{ijk} \mu_B B_j \hat{S}_k \tag{3}$$

where $T_{ijk} = \partial g_{jk}/\partial E_i$ are components of the Stark effect tensor regarding different directions. The $D_2$ local symmetry of $Ce^{3+}$ will eliminate most of them and merge the others into three, giving

$$\hat{H}_E = \mu_B \left[ E_x T_{xyz}\left(B_y \hat{S}_z + B_z \hat{S}_y\right) + E_y T_{yxz}\left(B_x \hat{S}_z + B_z \hat{S}_x\right) + E_z T_{zxy}\left(B_x \hat{S}_y + B_y \hat{S}_x\right) \right] \tag{4}$$

where parameters $T_{xyz}$, $T_{yxz}$ and $T_{zxy}$ can be determined by detecting the phase evolution with different applied $E$ fields, as detailed in SI. In our work, five different amplitudes from 0 to $10^6$ V/m were applied to the Ce:YAG single crystal. It can be clearly seen that the spin echo phase evolution frequency showed a linear response to the external $E$ field. It is worth noting that the rising and falling edges of the $E$ field pulse are about 40 ns. The induced magnetic field is at the mG level, comparable to the static and dynamic inhomogeneity from the magnet providing the $B_0$ field in the spectrometer, and thus neglected.

It is important to quantify the effect in order to optimize the manipulation speed of this quantum phase gate. The Stark effect shows strong anisotropy when the $E$ field pulse is applied in different directions in the $Ce^{3+}$ local coordinate system. By fitting the four experimental curves with calculated formula, the parameters are determined to be $T_{xyz} = 3.30(2) \times 10^{-8}$ m/V, $T_{yxz} = 8.76(6) \times 10^{-8}$ m/V and $T_{zxy} = 12.1(1) \times 10^{-8}$ m/V. This measurement, as well as all below, was conducted at 10 K.

With this set of parameters, we are able to optimize the efficiency of the $E$ field used to manipulate the spin quantum phase. Since the Stark effect is highly anisotropic, we can find the directions of the $E$ field and the external magnetic field ($B_0$) field in the $Ce^{3+}$ local coordinates that give the largest coupling strength. Through simulation, a coupling constant of 1.6 MHz·m/(T·MV) is acquired when the $B_0$ field is along (0.44, 0.90, 0) and the $E$ field is along (0, 0, 1) in the $Ce^{3+}$ local coordinates. With $E = 1$ MV/m and $B_0 = 0.6$ T, the coupling efficiency can be enhanced to 0.96 MHz with proper orientation.

Different from the effect of crystal-field and hyperfine-coupling terms, the energy shift induced by the $E$ field, originating from the Zeeman term, is also proportional to the $B_0$ field, as in Eq.3. Therefore, the manipulation efficiency of this $E$ field pulse can be enhanced by an increasing the $B_0$ field. A strong $B_0$ field provides plenty of other advantages, e.g., resolution enhancement, a high-quality initialized state, and a long phase memory time[24]. Conclusively, in the present research, it is safe to work with a relatively weak $E$ field.

**Electric field as a Phase Gate**

The above Stark effect experiment highlight some important information about the quantum state manipulation of Ce:YAG. This controllable effect could lead to an effective phase evolution. By applying the $E$ field pulse between the two mw pulses, the spin echo could gain an additional phase factor, which is controlled by both the $E$ field strength and the duration. Normally, mw pulses for electron spin manipulations are tens of nanoseconds in length. In our experiments, by optimizing the operating conditions of the $E$ field pulse, a $\pi/2$ phase evolution can be shortened to within 260 ns. In contrast, the phase memory time of Ce:YAG at 10 K is measured to be up to 15 μs. This results from the relatively high symmetry and the almost spin-free surrounding oxygen nuclei. A useful characterization of a qubit is its maximal number of effective operations before decoherence. In this research, a sequence of up to 57 $\pi/2$ rotations is allowed before the system decoheres. In quantum information processing, an $E$ field pulse can be considered as a quantum phase gate

$$\hat{R}(\phi) = e^{-i\hat{H}_E t} = \begin{pmatrix} e^{-i\phi/2} & 0 \\ 0 & e^{i\phi/2} \end{pmatrix} \tag{5}$$

which offers an additional phase difference $e^{i\phi}$.

This efficient quantum phase evolution offers us the possibility to demonstrate several quantum manipulations by the ***E*** field. Due to the higher operating rate, we can use the mw pulse to reverse the phase factor while the electron spin is undergoing the evolution in the external ***E*** field. In this demonstration, the electron spin is firstly prepared into ~~to~~ a superposition state in the ***xy*** plane of the Bloch sphere by a mw pulse. Then the ***E*** field pulse could drive a phase evolution of this superposition state, as if there were an additional magnetic field along the ***z*** axis. During this evolution driven by the ***E*** field, the system can be "kicked" by short mw π pulses to another state within the ***xy*** plane. Based on the pulse sequence used in the former experiment in Stark effect detection, these microwave π pulses were added symmetrically about the π pulse in the Hahn-echo sequence. During the ***E*** field pulse, a series of mw π pulses at arbitrarily chosen moments were applied to reverse the evolution of the spin polarization, as shown in Figure 3(a). This kind of operation is similar to the experiments with fullerene qubits[25]. The spin polarization evolves in the ***xy*** plane under the ***E*** field, while a mw π pulse drives the spin rapidly around the ***B*~1~** field along a half-circle, as illustrated in Figure 3(b).

In the above demonstration, the mw pulses operate much faster than the ***E*** field pulses. In the original proposal of bang-bang control, rapid spin flips are employed by short mw pulses to prevent the unwanted phase evolution[26]. As in Figure 3(c), when the system evolves in the ***xy*** plane driven by the ***E*** field and a series of mw π pulses are applied, one can see that the spin evolution is locked, and the spin is decoupled from the ***E*** field during the mw pulses. This is an example of the quantum Zeno effect, in which the quantum phase evolution is supressed by repeated measurements or controllable interactions with the environment[27]. The spin evolution is released at the end of each mw pulse, which can be locked again by the next. It is necessary to mention that, unlike the projective quantum Zeno effect experiment which locks the system in the eigenstate[28], this experiment can lock the system in any state.

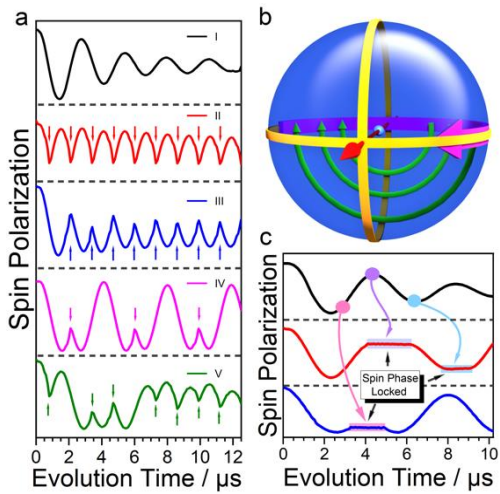

**Figure 3** The ***E*** field phase gate demonstration of quantum bang-bang control and the dynamic decoupling realized by mw pulses. **a,** (I) The phase evolution within the ***xy*** plane of the Bloch sphere driven by the ***E*** field phase gate. (II) to (V), Four trains of mw pulses to kick the spin echo. The arrows indicate the moment that the mw pulses are applied. The quantum phase evolution reverses when applying the microwave pulses. **b,** The Bloch sphere demonstration of bang-bang control. The bold purple curve represents the spin phase evolution induced by the ***E*** field phase gate, and the thin green curves indicate the mw "kicking" operation. The straight red arrow is the ***B*$_1$** field direction. **c,** Successive mw pulses, which were continuously implemented during the plateaus, can lock and release the phase evolution at arbitrary positions. The experiment was conducted at 10 K, with the microwave frequency 9.667 GHz.

**Phase gate application: D-J algorithm**

This efficient quantum phase gate could also help to demonstrate the refined D-J algorithm. The D-J problem is to determine whether a given oracle function $f:\{0,1\}^n \to \{0,1\}$, which takes an $n$-digit binary input and produces 0 or 1 as the output, is a balanced function (0 for half of the inputs and 1 for the other half) or a constant function (uniformly 0 or 1 for all inputs). A classical computer requires $2^{n-1}+1$ times of evaluation to solve this problem for the $n$-bit input situation. While quantum computing can solve it in a single try. This has been achieved with $^{19}$F nuclear spins[29], but not yet with electron spins. A refined version of the D-J algorithm proposed by Collins uses one of the input qubits as the output, allowing the $n$-bit problem to be solved using $n$ qubits[30]. Two types of quantum phase gates are required in our demonstration of the refined D-J algorithm with $n=1$. The first one, the Hadamard gate regarding the basis set $|s_z = \pm 1/2\rangle$, which can be performed by a $\pi/2$ mw pulse, is used to generate a uniform superposition of the two possible inputs and to convert the output to a readable eigenstate in the end. The second one, the $f$-controlled control gates are used to encode the functions for calculation. Here, the algorithm can be demonstrated in this two-level system with the $\boldsymbol{E}$ field quantum phase gate with a field strength of $0.8 \times 10^6$ V/m.

There are only four possible oracle functions for the single input situation in the D-J problem: $f_1(x) = 0$, $f_2(x) = 1$, $f_3(x) = 1-x$ and $f_4(x) = x$. These four functions can be encoded as 4 $f$-controlled gates realized by: $U_{f_1} = \begin{pmatrix} 1 & 0 \\ 0 & 1 \end{pmatrix} = \hat{R}(0)$, $U_{f_2} = \begin{pmatrix} -1 & 0 \\ 0 & -1 \end{pmatrix} = \hat{R}(2\pi)$, $U_{f_3} = \begin{pmatrix} -1 & 0 \\ 0 & 1 \end{pmatrix} = \hat{R}(\pi)$ and $U_{f_4} = \begin{pmatrix} 1 & 0 \\ 0 & -1 \end{pmatrix} = \hat{R}(-\pi)$. The superposition state $\frac{1}{\sqrt{2}}(|0\rangle + |1\rangle)$ generated by the first Hadamard gate is transformed to $\frac{1}{\sqrt{2}}[(-1)^{f_i(0)}|0\rangle + (-1)^{f_i(1)}|1\rangle]$ by $U_{f_i}$, and then transformed to $|0\rangle$ or $|1\rangle$ by the $\pi$ pulse and the second Hadamard gate for readout. The final Hahn echo is detected as the result, which can be $|0\rangle$ [for constant functions, $f_1$ and $f_2$, figure 4(b) and 4(c)] or $|1\rangle$ [for balanced functions, $f_3$ and $f_4$, figure 4(d) and 4(e)]. Unlike the previously reported implementation of the D-J algorithm in the diamond NV-centre[31], for which the phase shift was achieved by an auxiliary state via a $2\pi$ rotation, in this research, the phase shift gate is realized by the electric phase gate, and only a two-level system is necessary.

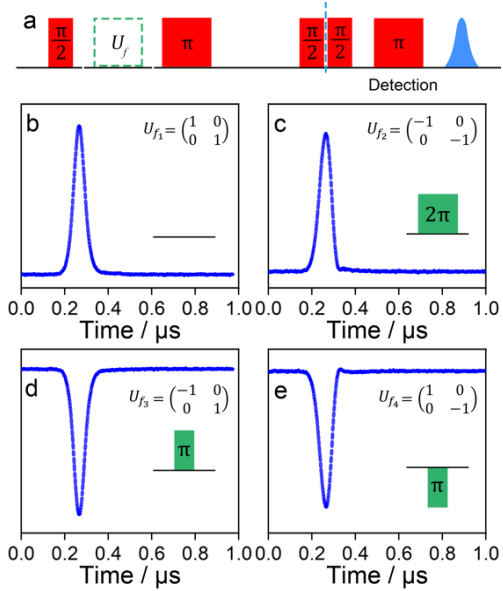

**Figure 4** The refined D-J algorithm ($n = 1$) using the $\boldsymbol{E}$ field pulses as the $f$-controlled gates. **a,** The pulse sequence of the D-J algorithm demonstration. The initial $\pi/2$ pulse is applied to prepare the superposition state of $\frac{1}{\sqrt{2}}(|0\rangle + |1\rangle)$. The orange block indicates the four $f$-controlled gates, as shown in the insets of **b** to **e**. The following $\pi/2$ pulse is to convert the evolved superposition state into eigenstates $|0\rangle$ (for constant functions, $f_1$ and $f_2$) or $|1\rangle$ (for balanced functions, $f_3$ and $f_4$).

The final Hahn echo pulse sequence is to read out the eigenstates |0⟩ or |1⟩. The spin echo results are shown in **b** to **e**, for which **b** and **c** with positive phases correspond to eigenstate |0⟩, whereas **d** and **e** with negative phases indicate eigenstate |1⟩. The experiment was conducted at 10 K, with the microwave frequency 9.667 GHz.

## Conclusion

We have demonstrated that an ***E*** field pulse applied to the Ce:YAG can be a highly efficient quantum phase gate. The phase gate is utilized to perform quantum bang-bang control and the refined Deutsch-Jozsa algorithm. The example of $Ce^{3+}$ with $4f^1$ configuration showed that rare-earth ions, with the aid of a proper symmetry, can enhance the spin-electric coupling via their strong spin-orbit interaction. The optimized manipulation condition with the X-band frequency and $10^6$ V/m field strength allows over fifty π/2 phase gate operations within the phase memory time of up to 15 μs at 10 K. The application of ***E*** field quantum phase gates enables us to achieve effective manipulations on the rare-earth qubit.

We would like to highlight the importance of spin-orbit coupling in the research on the ***E*** field coupling with the spin centre. The manipulation in our experiments, due to the strong spin-orbit coupling of the rare-earth ion, are much more efficient compared to those with other separated electron spin centres. The quantum coherence time, of course, is still one of the important factors that will determine if any application is possible, but faster manipulations allow us to be less picky about it. The ***E*** field in our experiments cannot exceed $10^6$ V/m due to the sample size, setting limit to the maximal number of manipulations before decoherence. The 1 MHz coupling constant might be enhanced further in other setups such as single spin break-junction devices, where the ***E*** field can easily reach $10^8$ V/m with a few volts. Such an ***E*** field is focused enough to control the electron spin individually.

Shortly after an earlier version of our manuscript was submitted to arXiv[32], a work by Liu et al. investigated into the ***E*** field effect on the quantum coherence involving a clock transition in the $HoW_{10}$ molecular nanomagnet, which provides a platform with even longer phase memory time for operation[33]. With other advantages of rare-earth ions such as ultralong optical coherence times, optical readout[34,35], and the possibility of device fabrication by chemical modifications and assembly, the high-efficiency quantum phase gate achieved with enhanced spin-electric coupling in this research indicates that electron spin qubits based on rare-earth ions make promising candidates for quantum information processing towards application.

## Author contributions

Z. L. performed the measurements, assisted by Y.-X. W. and Y.-H. F.. Z. L. and S.-D. J. processed the data and performed theoretical analysis. S.-X. Q. provided the methods to simulate the spin evolutions. Z.M.W. performed the single crystal face index. The project was conceived by S. G. and S.-D. J.. Z. L. and S.-D. J. designed the experiments and wrote the manuscript, assisted by Y.-X. W.. All the authors revised the manuscript.

## Conflicts of interest

The authors declare no conflict of interest.

## Keywords

Quantum coherent manipulation, spin-orbit coupling, spin-electric coupling, rare-earth ions, qubits.

## Teaser text

We utilized the strong spin-orbit coupling in a Ce:YAG single crystal to demonstrate efficient coherent manipulation with the electric field on this electron spin qubit.

**Acknowledgment**

This research is supported by the National Basic Research Program of China (2018YFA0306003 and 2017YFA0204903), National Natural Science Foundation of China (21822301, 11805024 and 11847301), Beijing Academy of Quantum Information Sciences (Y18G23).